\documentclass[aps,twocolumn,letterpaper,showpacs]{revtex4}
\usepackage{graphics,graphicx}
\usepackage{color}
\usepackage{wrapfig}
\usepackage{enumerate}
\usepackage{amssymb,amsmath}
\usepackage{bm}

\usepackage[normalem]{ulem} 

\begin{document}
\unitlength = 1mm	
\preprint{LA-UR-XX-XXXXX}
\title{Generalized spin-wave theory: application to the bilinear-biquadratic model}

\author{Rodrigo~A.~Muniz$^{1,2}$, Yasuyuki~Kato$^{3,4}$ and Cristian~D.~Batista$^3$}
\affiliation{$^1$Department of Physics, University of Toronto, Toronto, ON M5S 1A7, Canada \\ 
$^2$International Institute of Physics - UFRN, Natal, RN 59078-400, Brazil \\ 
$^3$Theoretical Division, T-4 and CNLS, Los Alamos National Laboratory, Los Alamos, NM 87545 \\
$^4$Riken CEMS, Wako-shi, Saitama 351-0198, Japan.}

\date{\today}
%
\begin{abstract}
We present a generalized spin-wave theory (GSWT) for treating spin Hamiltonians of arbitrary spin $S$.
The generalization consists of an extension of the traditional spin-wave theory from SU($2$) to SU($N$). 
Low energy excitations are waves of the local order parameter that fluctuates in the SU($N$) space of unitary transformations of the local spin states, instead of the SU($2$) space of local  spin rotations. 
Since the generators of the SU($N$) group  can be represented as bilinear forms in $N$-flavored bosons, the low-energy modes of the GSWT are described with $N-1$ different bosons.  
The generalization allows treating quantum spin systems whose ground state exhibit multipolar ordering as well as  detecting instabilities of magnetically ordered states (dipolar ordering) towards higher  multipolar orderings. 
We illustrate these advantages by applying the GSWT to a bilinear-biquadratic model of arbitrary spin $S$ on hypercubic lattices.
\end{abstract}
\maketitle
\section{Introduction}
Spin-wave theory (SWT) is the simplest and most popular approach for treating quantum spins systems that have a magnetically
ordered ground state and small quantum fluctuations. For systems with bilinear spin interactions, such as exchange or dipole-dipole interactions,
the relative strength of the quantum fluctuations scales as $1/\sqrt{S}$, implying that the SWT becomes increasingly accurate for larger values 
of the spin $S$. This is the reason why the SWT is regarded as a semi-classical approach. The only type of ground state ordering that survives in the classical
limit,  in which spins are replaced by a classical vector field, is the usual dipolar ordering. Fluctuations  of the classical vector field are entirely described by
the SO(3) group of local rotations, which is isomorphic to SU(2)/Z$_2$, and  those are the only fluctuations which are included in the traditional SWT. 

It is well known that the spectrum of possible ground state orderings of quantum spin systems is much richer than the usual dipolar ordering: $\langle {\bm S}_{\bm r}  \rangle \neq 0$. Nematic, octupolar or any higher multipolar orderings ($\langle {\bm S}_{\bm r}  \rangle =0$) also emerge  under the right conditions. Because such ground state orderings
do not have a classical counterpart at zero temperature ($T=0$), it is clear that the SWT has to be generalized in some way. Moreover, even if the ground state exhibits the usual dipolar magnetic ordering, the local fluctuations of a spin $S$ around a particular mean field state are described by the SU($2S+1$) group of local unitary transformations, instead of the SU(2) group of rotations. This observation indicates that a generalization of the traditional SWT can also be useful for describing the excitations of spin systems whose ground state exhibits dipolar magnetic ordering.

In this paper we introduce a generalization of the SWT that allows for describing the low-energy modes of spin systems with arbitrary orderings (spontaneous 
or induced by external fields). The generalization simply consists of extending the SU(2) group of local spin rotations to the SU($N$) group of 
local unitary transformations, with $N$ being equal to the dimension of the local Hilbert space. We show that the generalized spin wave theory (GSWT)  is also more appropriate for describing spin systems that exhibit the usual dipolar magnetic ordering  ($\langle {\bm S}_{\bm r}  \rangle \neq 0$), but are close to a multipolar instability. Multipolar orderings are quite common in $f$-electron materials because of the strong single-ion anisotropies that result from  the
dominant spin-orbit coupling (the effective magnetic moments can have large orbital contributions). For instance, antiferroquadrupolar ordering has been observed in CeB$_6$\cite{shiina1997,sakai1997}, while  CeAg\cite{ray1976} exhibits ferroquadrupolar ordering. Quadrupolar  ordering has also been proposed for NiGa$_2$S$_4$ \cite{tsunetsugu2006,lauchli2006}.

To illustrate the advantages of using the GSWT in comparison to the standard SWT, we consider a simple bilinear-biquadratic spin model for arbitrary spin $S$ on hypercubic lattices and compare the spectra of low-energy modes obtained for phases with usual dipolar ordering. 
Indeed, Papanicolaou applied the GSWT to the bilinear-biquadratic $S=1$ model and obtained reasonable predictions \cite{papanicolaou1988}. The main purpose of our paper is to provide a mathematical framework for the GSWT  that unveils the physical situations for which the method is much more appropriate than the ordinary SWT. 

%

\section{Generalized spin-wave theory}
In this section, we consider an arbitrary quantum spin Hamiltonian  defined on some lattice. 
The lattice can be decomposed into a convenient set of units containing a finite number of degrees of freedom. 
For instance, the unit may be a single site, a dimer or a trimer. 
We will assume that the local Hilbert space of a unit has dimension $N$, e.g. , $N=2S+1$ if the unit is a single spin $S$.
We will introduce a set of generators of SU($N$) and use the local Hilbert space, ${\cal H}_{\bm r}$,
as the vector space of the fundamental representation of the SU($N$) group. 
For this purpose we consider a set of $N$ Schwinger bosons (SBs) $b^{\;}_{{\bm r} m}$ ($0 \leq m \leq N-1$) \cite{auerbach98}, that satisfy the local constraint 
\begin{equation}
	\sum_{m=0}^{N-1} b^{\dagger}_{{\bm r} m} b^{\;}_{{\bm r} m} = N {\cal S},
	\label{eq:constr}
\end{equation}
and provide a simple way of representing the SU($N$) generators. 
The local constraint of Eq.~\eqref{eq:constr} projects the bosonic operators into the physical space of dimension $N$.
The SU($N$) generators are bilinear forms in the SBs:
\begin{equation}
	O^{m m'} = b^{\dagger}_{{\bm r} m} b^{\;}_{{\bm r} m'},
	\label{sun}
\end{equation}
that satisfy the SU($N$) commutation relations
\begin{equation}
[ O^{mm'},O^{nn'} ] = \delta_{m'n} O^{mn'} - \delta_{mn'} O^{nm'}.
\label{comm}
\end{equation}

The value of $N {\cal S}$ depends on $S$, which sets the representation of SU($N$). For instance, if $S=\frac{5}{2}$ and $N=3$ then $N {\cal S}=2$, as the second representation of SU($3$) has dimension $6=2S+1$. In this case it is possible to describe quadrupolar  fluctuations of the order parameter on a SU($3$) subspace of the spin $S=\frac{5}{2}$ states, with a Hamiltonian of only quadratic order on SBs. But octupolar(or higher multipolar) excitations would only be contained in higher order expansions of the Hamiltonian, which would require dealing with interacting quasiparticles.

In contrast, if $N=2S+1$ and $N {\cal S}=1$, which corresponds to the fundamental representation of SU($N$), all  the local fluctuations of the spin field are described by a quadratic order expansion of the Hamiltonian in terms the SBs  $b^{\dagger}_{{\bm r} m}$, $b^{\;}_{{\bm r} m}$.    The identity and the $N^2-1$ generators of SU($N$) form a basis for the $N \times N$ matrices acting on the linear space ${\cal H}_{\bm r}$. Therefore, any local operator can be expressed as a linear 
combination of the operators $O^{m m'} $: a local operator $X_{\bm r}$ is represented as
\begin{equation}
 X_{\bm r} = \sum_{m,m'}  {\cal X}_{\bm r}^{m m'}  O^{m m'}_{\bm r} = {\bf b}^{\dagger}_{{\bm r}} {\cal X}_{\bm r} {\bf b}^{\;}_{{\bm r}},
 \label{linex}
\end{equation}
with 
${\bm b}^{\dagger}_{\bm r} \equiv
	(
	b^{\dagger}_{ {\bm r} 0} , b^{\dagger}_{ {\bm r} 1}, \cdots,
	b^{\dagger}_{{\bm r} N-1}
	) $.
Eq.~\eqref{linex} shows that $X_{\bm r} $  is a linear combination of  the identity and the  $N^2-1$ components of the local SU($N$) order parameter. 
If we are considering single-spin units, 
the $N^2-1$ components of the local SU($N$) order parameter can be decomposed in the different  irreducible representations of SU(2) 
which are obtained from tensor products of the spin operators $S^{\nu}_{\bm r}$. 
For instance, $S=1$ spins  in the $N=3$, $N {\cal S}=1$ representation include the local (dipolar) magnetic moment, ${S}^{\nu}_{\bm r}$,  
and nematic (quadrupolar) moment, $Q^{\mu \nu}_{\bm r}= S^{\mu}_{\bm r} S^{\nu}_{\bm r} - 2 \delta_{\mu \nu}/3$, with dimension 3 and 5, respectively. 
The sum 3+5=8 coincides with the number of generators of SU(3) or the number of components of the local SU(3) order parameter \cite{Batista02,Batista04}. 
$S=3/2$ spins additionally admit an octupolar local order parameters that are obtained from tensor products of three spin operators and has dimension 7. 
Note that 3+5+7=15 is the number of generators of SU(4). 

%
%
%

It is clear by now that an arbitrary local order parameter is always contained in the most general SU($N$) order parameter of $N^2-1$ components. 
Consequently, it is necessary and sufficient to consider this SU($N$) order parameter instead of  the local magnetization ${\bm S}_{\bm r}$
for constructing the most general mean-field theory.
Therefore, the waves of the generalized ``spin-wave" theory are not only magnons. In general, they are waves of the underlying ground state multipolar ordering (e.g.,  quadrupolar waves or octupolar waves). We note, however, that waves of a non-dipolar order parameter, like nematic or octupolar, should always include a small dipolar component because the order parameter is  fluctuating in the SU($N$) order parameter space. In other words, if the mean value of the order parameter is pointing along a non-dipolar direction, small fluctuations produced by unitary transformations, {\it which are different from 
spin rotations}, will induce a small dipolar component.  This general observation has important consequences for the experimental observation of
the ``hidden'' multipolar orderings. For instance, Smerald {\it et al} recently proposed to detect nematic ordering by detecting the small dipolar component of the nematic waves with inelastic neutron scattering (INS) or nuclear magnetic resonance (NMR) \cite{Smerald13}. We note that the same logic can be applied to any other multipolar spin ordering.

%
%
The previous analysis already shows the advantages of working with the SU($N$) space of unitary transformations instead of restricting to
the SU(2) space of local spin rotations. We will consider now  the simplest case of usual magnetic ordering to emphasize this point even more. 
The low-energy modes of the dipolar ordering are magnons. However, if the system approaches a nematic critical point, nematic-waves that 
are bound states of two magnons, eventually become part of the spectrum of low-energy excitations. To capture this effect with the ordinary 
SWT it is necessary to go beyond the linear approximation and solve the two magnon problem. In contrast, this effect is captured by the  
GSWT already at the linear level. Whereas higher order multipolar instabilities  are also captured by the linear GSWT, an ordinary SWT treatment
would require to consider  $n$-magnon processes ({\it e.g.}, $n=2$ and $n=3$ for nematic and octupolar orderings respectively). 

SWT is based upon the Holstein-Primakoff (HP) bosons that provide a useful representation of the generators of SU(2). 
Therefore, a natural generalization of the SWT is obtained by extending the HP representation from SU(2) to SU($N$). 
This generalization can be done by  condensing one of the $N$ SBs: the corresponding creation and annihilation operators are replaced by a number according to  the constraint of Eq.~\eqref{eq:constr}. The SB that is condensed is the one which creates the local state that minimizes the mean field energy, i.e., 
the mean value of the Hamiltonian over the variational space of direct product sates
\begin{equation}
| \psi_{\rm mf} \rangle = \prod_{\bm r} \tilde{b}^{\dagger}_{{\bm r} 0} | \emptyset \rangle.
\label{min}
\end{equation}
In general, this $\tilde{b}^{\dagger}_{{\bm r} 0} $ is a linear combination of the original $N$ SBs that create a particular basis of ${\cal H}_{\bm r}$. Therefore, it is necessary 
to make a unitary transformation, $\tilde{\bm{b}}_{\bm r}= U \bm{b}_{\bm r}$, that maps the original SB's into a new set, whose
$m=0$ boson is the one to be condensed:

\begin{equation}
\tilde{b}^{\dagger}_{ {\bm r} 0}  = \tilde{b}^{\;}_{ {\bm r} 0} = \sqrt{N {\cal S} } \sqrt{1- \frac{1}{N {\cal S} } \sum_{m=1}^{N-1} \tilde{b}^{\dagger}_{{\bm r} m} \tilde{b}^{}_{{\bm r} m}}.
\label{cond}
\end{equation}  
This transformation corresponds to choosing the  quantization axis along the direction of the local SU($N$) order parameter, as it is done in the traditional SWT
with the SU(2) HP bosons.
The HP representation of the SU($N$) generators is given by $ O^{m m'} = {\tilde b}^{\dagger}_{{\bm r} m} {\tilde b}^{\;}_{{\bm r} m'}$, where  
$\tilde{b}^{\dagger}_{ {\bm r} 0}$ $ \tilde{b}^{\;}_{ {\bm r} 0} $ have to be replaced by the expression given in Eq.~\eqref{cond}.
The approximation 
$ \tilde{b}^{\dagger}_{ {\bm r} 0}  = \tilde{b}^{\;}_{ {\bm r} 0} \simeq \sqrt{N {\cal S}} \left( 1- \frac{1}{2 N {\cal S}} \underset{m=1}{\sum} \tilde{b}^{\dagger}_{{\bm r} m} \tilde{b}^{}_{{\bm r} m} \right) $ 
is justified by assuming that only a few bosons are not part of the condensate 
 $\underset{m=1}{\sum} \langle \tilde{b}^{\dagger}_{{\bm r} m} \tilde{b}^{}_{{\bm r} m} \rangle  \ll N {\cal S}$. 
After making this approximation, the  expression for a general operator becomes 
\begin{equation}
\begin{array}{rl}
X_{\bm r} = &
N {\cal S} \tilde{\cal X}^{00}_{\bm r} + \sqrt{N {\cal S}} \underset{m=1}{\sum} 
\left( \tilde{b}_{{\bm r} m}^{\dagger} \tilde{\cal X}^{m0}_{\bm r} + \tilde{\cal X}^{0m}_{\bm r} \tilde{b}_{{\bm r} m} \right) -
\\
 & - \tilde{\cal X}^{00}_{\bm r} \underset{m=1}{\sum} \tilde{b}_{{\bm r} m}^{\dagger} \tilde{b}_{{\bm r} m} + \underset{m m'}{\sum} \tilde{b}_{{\bm r} m}^{\dagger} \tilde{\cal X}^{m m'}_{\bm r} \tilde{b}_{{\bm r} m'}.
\end{array}
\end{equation} 
The $N-1$ non-condensed SBs become the SU($N$) HP bosons for our spins. 
The SU($N$) generators written as quadratic operators in the HP representation still satisfy the SU($N$) commutation relations of Eq.~\eqref{comm}.

The rest of the procedure is rather straightforward. 
The spin Hamiltonian under consideration is written in the SU($N$) HP representation, and 
we only keep terms up to quadratic order in the bosonic operators. 
The result is a quadratic  Hamiltonian (the linear terms cancel up after the minimization condition  of Eq.~\eqref{min}) 
that is diagonalized by means of a standard  Bogoliubov transformation. 
%

\section{Bilinear-biquadratic model}

The simplest examples of non-dipolar orderings are provided by $S=1$ systems that can exhibit nematic or quadrupolar ground state 
orderings which are either induced by an external field, such as the crystal field  \cite{Zapf06}, or the result of a spontaneous symmetry breaking
\cite{penc2011,wierschem2012}. For instance, ferroquadrupolar spin ordering was originally  proposed as one of the ordered phases of the spin one bilinear biquadratic model \cite{blume1969}. The existence of the ferroquadrupolar phase was confirmed for the spin one bilinear biquadratic model defined on a square lattice  by applying  unbiased quantum Monte-Carlo simulations~\cite{harada2002}. To illustrate the advantages of using the GSWT, we will consider the same  model for arbitrary spin $S$ and hypercubic lattices, written in terms of generators of the SU($N=2S+1$) group [SU($N=2S+1$) spins] in the fundamental representation: $N {\cal S}=1$ . 
The corresponding spin Hamiltonian is
\begin{eqnarray}
	\mathcal{H} &=& J_L\sum_{\langle {\bm r}, {\bm r}'\rangle} {\bm S}_{\bm r} \cdot {\bm S}_{\bm r'}
	+J_Q\sum_{\langle {\bm r}, {\bm r}'\rangle} \left({\bm S}_{\bm r} \cdot {\bm S}_{\bm r'}\right)^2,
	\label{eq:hamiltonian}
\end{eqnarray}
where the summations run over all nearest neighbor sites.
For convenience, we introduce the angle $\alpha$, $J_L=J\cos\alpha$, and $J_Q=JS^{-2}\sin\alpha$, to parametrize the family of bilinear biquadratic
Hamiltonians. The $S^{-2}$ factor is introduced in the parametrization of $J_Q$ to make the bilinear and biquadratic terms comparable in the large-$S$ limit.
We will consider both positive and negative values of the exchange coupling $J_L$.
It is well know that the ground state exhibits antiferromagnetic (AFM) ordering for positive values of $J_L$ and small enough values of $|J_Q|$.
To consider the  two sublattice AFM ordering, it is convenient to rotate the spin reference frame of the $B$ sublattice along $x$-axis by an angle $\pi$: 
$S^{y,z} \to -S^{y,z}$, and $S^x\to S^x$. This unitary transformation maps an N{\'e}el state polarized along the $z$-axis into a FM state. 
For a general expression of the Hamiltonian, we introduce ${\bm a}=(1,1,1)$ for the ferromagnet and 
$(1,-1,-1)$ for antiferromagnet.
In the new basis, the Hamiltionian \eqref{eq:hamiltonian} becomes
\begin{eqnarray}
	\mathcal{H} &=& J_L \!\!\!\! \sum_{\langle {\bm r}, {\bm r}'\rangle,\nu} \!\!\!\! a_{\nu} {S}^{\nu}_{\bm r} {S}^{\nu}_{\bm r'}
	+J_Q \!\!\!\! \sum_{\langle {\bm r}, {\bm r}'\rangle,\nu,\mu}  \!\!\!\! a_{\nu} a_{\mu} {S}^{\nu}_{\bm r} {S}^{\mu}_{\bm r'} {S}^{\nu}_{\bm r}{S}^{\mu}_{\bm r'},
	\label{eq:hamiltonianR}
\end{eqnarray}
where $\nu$ and  $\mu$ run over $\{ x, y, z\}$.

\section{Stability of dipolar phases}
In this section we study the  instabilities of the usual ferromagnetic (FM) and AFM dipolar orderings, that occur when
$J_Q$ reaches certain critical values. Namely, we calculate the excitation spectra of the  FM and AFM phases
and identify the points where a  branch of excitations, that is gapped at $J_Q=0$, becomes gapless.
The nature of the excitation that becomes gapless tells us the kind of multipolar fluctuations that diverge at those  quantum critical points and therefore the type
of non-dipolar ordering that could become stable for slightly larger values of $|J_Q|$. 
Note, however, that this procedure is not applicable if the quantum phase transition is of  first order. Nevertheless, minimization 
over the variational space  of product or mean field states \eqref{min}  normally reveals the existence of a first order transition
between two ground states that break different symmetries.

The ground state ordering of  ${\cal H} (J_Q=0)$ is FM for $J_L<0$ ($\alpha = \pi$) and AFM for $J_L>0$ ($\alpha = 0$).
The mean field state for the FM ordering is the fully polarized state that satisfies  $S^z_{\bm r} | \psi_{\rm mf} \rangle =S | \psi_{\rm mf} \rangle$ for all ${\bm r}$.
The same mean field state describes the AFM ordering after performing the above mentioned unitary transformation, $S^{y,z} \to -S^{y,z}$, and $S^x\to S^x$,
that maps  ${\bm a}=(1,1,1)$ into ${\bm a}=(1,-1,-1)$.

The eigenstates of $S^z_{\bm r}$, $\left| S^z_{\bm r} \right\rangle = b^{\dag}_{{\bm r}S-S^z} \left| \emptyset \right\rangle$,  are a convenient choice of basis for the  SU($N=2S+1$) SBs. The $b^{\dag}_{{\bm r} 0 }$ boson already creates the fully polarized state that minimizes the mean field energy, so we do not need 
to perform and additional unitary transformation ($U=I$). The Hamiltonian ${\cal H}$ can now be expressed in terms of  these SBs by using using Eq.~\eqref{linex}:
\begin{eqnarray}
\mathcal{H} &=& J_L\sum_{\langle {\bm r}, {\bm r}'\rangle,\nu} a_{\nu}
	 {\bm b}^{\dag}_{\bm r} \mathcal{S}^{\nu} {\bm b}^{\;}_{\bm r} 
	 {\bm b}^{\dag}_{\bm r'} \mathcal{S}^{\nu} {\bm b}^{\;}_{\bm r'}
	 \nonumber \\
	&+& J_Q\sum_{\langle {\bm r}, {\bm r}'\rangle,\nu,\mu} a_{\nu} a_{\mu} 
	 {\bm b}^{\dag}_{\bm r} \mathcal{S}^{\nu\mu} {\bm b}^{\;}_{\bm r} 
	 {\bm b}^{\dag}_{\bm r'} \mathcal{S}^{\nu\mu} {\bm b}^{\;}_{\bm r'},
\label{eq:HF}
\end{eqnarray}
where
\begin{eqnarray}
\mathcal{S}^{x}_{mm'}&=& \delta_{m\; m'-1} \frac{ \sqrt{(m+1)(2S-m)} }{2}
\nonumber \\
&+& \delta_{m-1\; m'} \frac{ \sqrt{(m'+1)(2S-m')} }{2} ,
\nonumber\\
\mathcal{S}^{y}_{mm'}&=& \delta_{m\; m'-1} \frac{ \sqrt{(m+1)(2S-m)} }{2i} 
\nonumber \\
&-& \delta_{m-1\; m'} \frac{ \sqrt{(m'+1)(2S-m')} }{2i} ,
\nonumber\\
\mathcal{S}^{z}_{mm'}&=& \delta_{m\;m'} (S - m),
\end{eqnarray}
and
\begin{eqnarray}
\mathcal{S}^{\nu\mu}_{mm'}&=&
\sum_{m''} 
\mathcal{S}^{\nu}_{mm''}
\mathcal{S}^{\mu}_{m''m'}.
\nonumber
\end{eqnarray}
Here, $\mathcal{S}^{\nu}$ is the matrix associated with the local spin operator $S^{\nu}_{\bm r}$, while $\mathcal{S}^{\nu\mu}$ is the matrix associated with 
local bilinear operator $S^{\nu}_{\bm r}  S^{\mu}_{\bm r}$.

The next step is to perform the Holstein-Primakoff transformation \eqref{cond}, 
\begin{equation}
{b}^{\dagger}_{ {\bm r} 0}  = {b}^{\;}_{ {\bm r} 0} = \sqrt{1- \sum_{m=1}^{N-1} {b}^{\dagger}_{{\bm r} m} {b}_{{\bm r} m}},
\end{equation}  
and keep the terms up to quadratic order in the HP bosonic operators:
\begin{eqnarray}
\mathcal{H}  &=& dN_{s} t^{00}_{00}+\mathcal{H}_{\rm GSW}+\cdots,
\end{eqnarray}
with
\begin{eqnarray}
\mathcal{H}_{\rm GSW} \!\!   &=& \!\!\!\! \!\!\!\! 
\sum_{\langle {\bm r}, {\bm r}'\rangle, m, m'} \!\!\!\! 
[
t^{m0}_{0m'} b^{\dag}_{{\bm r}m}b_{{\bm r}'m'}
+ t^{m0}_{m'0}b^{\dag}_{{\bm r}m}b^{\dag}_{{\bm r}'m'}
+{\rm H.c.}
]
\nonumber\\
&+& 
2d
\sum_{\langle {\bm r}, {\bm r}'\rangle, m, m'} \!\!\!\! 
\left( t^{mm'}_{00} -t^{00}_{00} \delta_{mm'}\right) b^{\dag}_{{\bm r}m} b_{{\bm r}m'},
\end{eqnarray}
and 
\begin{eqnarray}
t^{m_0m_1}_{m_2m_3}&=& J_L\left( \sum_{\nu} a_{\nu} \mathcal{S}^{\nu}_{m_0m_1} \mathcal{S}^{\nu}_{m_2m_3} \right)
\nonumber \\
&+&J_Q \left( \sum_{\nu,\mu} a_{\nu}  a_{\mu} \mathcal{S}^{\nu\mu}_{m_0m_1} \mathcal{S}^{\nu\mu}_{m_2m_3} \right).
\nonumber
\end{eqnarray}
$N_{s}$ is the number of lattice sites and $ 1 \leq m,m' \leq N-1$.
There is no linear contribution in the bosonic operators because we are expanding around the mean field state $| \psi_{\rm} \rangle $ that 
minimizes the energy.
By going to momentum space,
\begin{equation}
 b^{\dag}_{{\bm k}m}= L^{-1/2} \sum_{\bm r} b^{\dag}_{{\bm r}m}\exp[i{\bm k}\cdot{\bm r}],
 \end{equation}
where $L$ is the linear size of the system, we obtain
\begin{eqnarray}
\mathcal{H}_{\rm GSW} \!\! &=& \!\!
\sum_{{\bm k}, m,m'} \!\!\!\! 
\left[
t^{m0}_{0m'} \gamma^{+}_{\bm k} b^{\dag}_{{\bm k}m}b_{{\bm k}m'}
+t^{m0}_{m'0}  \gamma^{+}_{\bm k} b^{\dag}_{{\bm k}m}b^{\dag}_{-{\bm k}m'}
+{\rm H.c.}
\right]
\nonumber\\
&+& 
2d
\sum_{{\bm k}, m,m'} \!\!\!\! 
\left( t^{mm'}_{00} -t^{00}_{00} \delta_{mm'}\right) b^{\dag}_{{\bm k}m} b_{{\bm k}m'},
\label{hgsw}
\end{eqnarray}
with $\gamma^{+}_{\bm k}= \sum_{\eta=1}^{d} e^{ i k_{\eta}}.$

\subsection{Excitation spectrum and stability of ferromagnetic phase}

Because ${\bm a}=(1,1,1)$ for  the FM mean field ground state,  Eq.~\eqref{hgsw} leads to the following expression for $\mathcal{H}^{\rm FM}_{\rm GSW}$:
\begin{eqnarray}
\mathcal{H}^{\rm FM}_{\rm GSW} = \sum^{2S}_{m=1}  \epsilon^{f}_{{\bm k} m} b_{{\bm k}m}^{\dagger} b_{{\bm k}m},
\end{eqnarray}
with
\begin{eqnarray}
\label{gswdrf}
\epsilon^f_{{\bm k} 1} \!\!\! &=& \!\!\!  -JS\left[	\cos\alpha +2\left(1-\frac{1}{S}\right) \sin\alpha \right] \left(2d-\gamma_{\bm k}\right),
\nonumber\\
\epsilon^f_{{\bm k} 2} \!\!\! &=& \!\!\!	-4dJS\left[	\cos\alpha +\left(2 - \frac{4S-1}{S^{2}} \right) \sin\alpha\right]
\nonumber \\
&-& 2J\left(1-\frac{1}{2S}\right)\sin\alpha  \left(2d -\gamma_{\bm k} \right),
\end{eqnarray}
and
\begin{eqnarray}
\epsilon^f_{{\bm k} m} \!\! = \!\!  -  2mdJS\left[\cos\alpha+\left(2-
  \frac{2\left( m+1\right)S-m+1}{2S^{2}}\right)\sin\alpha \right], \nonumber 
\end{eqnarray}
for $ 3 \leq m \leq 2S$ where $\gamma_{\bm k}= 2\sum_{\eta=1}^{d} \cos k_{\eta}$.

As expected, the dispersion relation for the single branch of magnon excitations that is obtained with the ordinary SWT, 
\begin{eqnarray}
\omega_{\bm k} =-JS(\cos \alpha + 2 \sin\alpha)(2d-\gamma_{\bm k}),
\label{eq:wkfmswt}
\end{eqnarray}
is equal to  $\epsilon^f_{{\bm k} 1} $ for $\alpha=\pi$ because the $m=1$ bosons describe exactly the same single-magnon modes. However, both dispersions become different for $\alpha \neq \pi$ (finite biquadratic term) and they only coincide  in the $S\rightarrow \infty$ limit. This difference in the SWT and GSWT single-magnon dispersions leads to different stability ranges of the FM phase.
Within the SWT, the FM phase becomes unstable at $\cos \alpha + 2 \sin \alpha > 0$.
As we will see below, the GSWT predicts a quite different stability range that coincides with the phase diagram obtained from numerical and/or mean field treatments.

\begin{table*}
	\begin{tabular}{c|cc|cc}
	\toprule
	$S$&$\alpha_{\rm{min}}^{\rm F}$ &$m_{\rm soft}$& $\alpha_{\rm{max}}^{\rm{F}}$ &$m_{\rm soft}$    
	\\ \colrule
	1                  &$\frac{\pi}{2}$& 1 and 2 &$\frac{5\pi}{4}$& 2\\
	$\frac{3}{2}$&$\cos \alpha=  -\frac{2}{\sqrt{13}}\cap \sin \alpha= \frac{3}{\sqrt{13}} $& 1 and 2
	&$\cos \alpha=-\frac{2}{\sqrt{85}}  \cap \sin \alpha=-\frac{9}{\sqrt{85}}  $&2 and 3\\
	2&$\cos \alpha=  -\frac{1}{\sqrt{2}}\cap \sin \alpha= \frac{1}{\sqrt{2}} $&1 and 2
	&$\cos \alpha= -\frac{1}{\sqrt{65}}  \cap \sin \alpha=-\frac{8}{\sqrt{65}}  $&4\\
	$\frac{5}{2}$&$\cos \alpha= -\frac{6}{\sqrt{61}}  \cap \sin \alpha=  \frac{5}{\sqrt{61}} $&1 and 2
	&$\cos \alpha= -\frac{2}{\sqrt{629}} \cap \sin \alpha= -\frac{25}{\sqrt{629}} $& 5\\
	\botrule
	\end{tabular}
	\caption{Range of stability of the FM phase that is extracted from the analysis of the excitation spectrum predicted by the GSWT. }
	\label{tab1}
\end{table*}

\begin{figure*}[htpb]
 \includegraphics[angle=0,width=16cm]{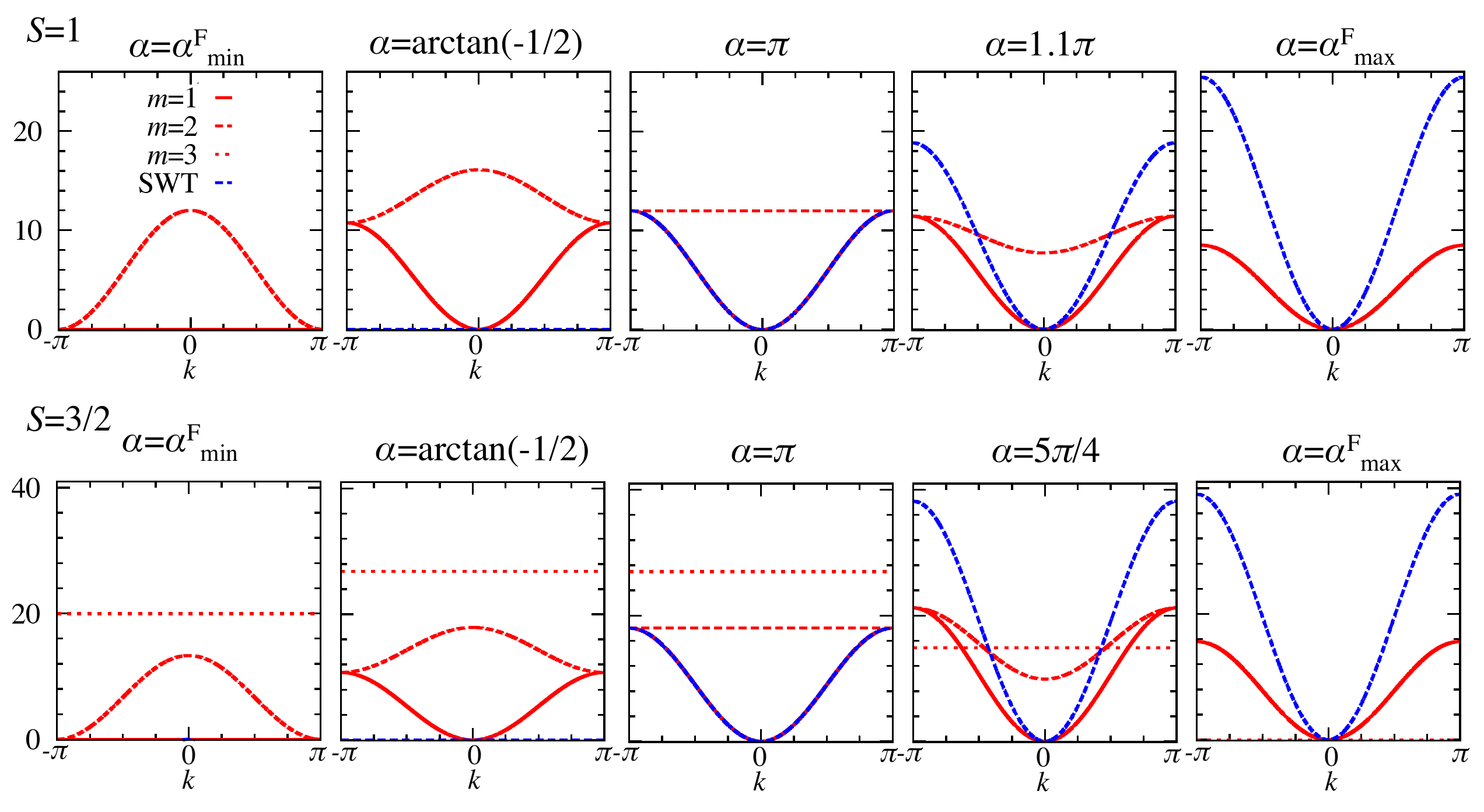}
 \caption{(Color online) 
 	Excitation spectra of the SWT and the GSWT for $S=1$ (upper panels) and $S=3/2$ (lower panels), $\alpha^{F}_{\rm min} \leq \alpha \leq  \alpha^{F}_{\rm max}$ and $d=3$. The label $k$ of the horizontal axes corresponds to a parametrization of the wave-vectors along the $(1,1,1)$ direction: ${\bm k}=(k,k,k)$.
 }
 \label{fig:dispF}
\end{figure*}

The stability conditions that arise in the GSWT by requiring that all the modes must have semi-positive defined frequencies  are:
\begin{eqnarray}
S \cos\alpha + 2\left(S - 1 \right) \sin\alpha & < & 0, \nonumber\\
S \cos\alpha + \left(2S - 3 + \frac{1}{2S} \right) \sin\alpha + \left( 1 - \frac{1}{2S} \right)\left|\sin \alpha \right|  & < &0,
\nonumber\\
S \cos\alpha + \left( 2S - m - 1 + \frac{m-1}{2S} \right) \sin\alpha & < &0,
\nonumber \\
\label{condfm}
\end{eqnarray}
for $ 3 \leq m \leq 2S$. 
Table~\ref{tab1} summarizes the expected phase boundaries and the modes that become soft for $1 \leq S \leq  5/2$.

Fig.~\ref{fig:dispF} shows the evolution of the dispersions of different modes that are obtained with the 
GSWT as well as the single-magnon dispersion  that is obtained with the ordinary SWT [see Eq.~\eqref{eq:wkfmswt}]. 
It is clear from the figure that this dispersion is quite different from the one obtained  for the $m=1$ mode with the GSWT. AS we already mentioned, both dispersions coincide 
only for the pure bilinear model ($\alpha=\pi$). Away from that special point, SWT predicts a magnon dispersion that is much flatter for
$\alpha< \pi$ and significantly broader for $\alpha >\pi$. Direct comparison of Eqs.~\eqref{gswdrf} and \eqref{eq:wkfmswt} shows that the difference 
between GSWT and SWT arises from the $(1-1/S)$ factor that does not appear in Eq.~\eqref{eq:wkfmswt}.  This $1/S$ correction is obviously important for low spin systems like $S=1$. The correction arises from biquadratic contributions of the form $S^z_{\bm r} S^z_{\bm r'} (S^+_{\bm r} S^-_{\bm r'}+ S^-_{\bm r} S^+_{\bm r'}) + (S^+_{\bm r} S^-_{\bm r'}+ S^-_{\bm r} S^+_{\bm r'}) S^z_{\bm r} S^z_{\bm r'}$. It is clear that these terms should not contribute to the single-magnon ($m=1$) dispersion of an $S=1$ ferromagnet because the local spin state that is obtained after flipping a single spin at site ${\bm r}$ is an $S^z_{\bm r}=0$ state. The operator $S^z_{\bm r} S^z_{\bm r'} (S^+_{\bm r} S^-_{\bm r'}+ S^-_{\bm r} S^+_{\bm r'}) + (S^+_{\bm r} S^-_{\bm r'}+ S^-_{\bm r} S^+_{\bm r'}) S^z_{\bm r} S^z_{\bm r'}$ is exactly equal to zero when projected into the subspace generated by these states. However, the operator $ \langle S^z_{\bm r} S^z_{\bm r'} \rangle (S^+_{\bm r} S^-_{\bm r'}+ S^-_{\bm r} S^+_{\bm r'}) + (S^+_{\bm r} S^-_{\bm r'}+ S^-_{\bm r} S^+_{\bm r'}) \langle S^z_{\bm r} S^z_{\bm r'} \rangle$ that appears in the linear SWT is finite. This shortcoming of the ordinary SWT has important consequences for relatively low-spin systems like the ones considered in Fig.~\ref{fig:dispF}. In particular, SWT predicts an instability at 
$\alpha={\rm atan}(-1/2)$ while it is known that the FM phase remains stable down to much lower values of $\alpha$. Indeed, the phase boundaries obtained from numerical calculations for $S=1$~\cite{toth2012} are  $\alpha^{F}_{\rm min} = -\pi/2$ and $\alpha^{F}_{\rm max}=\pi/4$ [SU(3) FM~\cite{Batista04}] in perfect agreement with the prediction of our GSWT (see Table~\ref{tab1}). As expected, SWT also fails to capture the antiferroquadrupolar instability at $\alpha = \alpha^{F}_{\rm max}=\pi/4$. The situation is rather similar for $S=3/2$. Although we are not aware of
the existence of numerical results for this case, direct minimization of $\langle \psi_{\rm mf} | {\cal H} |\psi_{\rm mf} \rangle$ 
(see Eq.~\ref{min}) over two and three-sublattice structures leads to  {\it exactly the same stability range $\alpha^{F}_{\rm min} < \alpha < \alpha^{F}_{\rm max}$ obtained by analysing the low-energy modes of the GSWT}  (Eq.~\ref{condfm}). This mean field 
stability range of the FM phase is significantly bigger than the one predicted by ordinary SWT. 

The simple message of the previous discussion is that the GSWT not only includes $m>1$ low-energy modes that become important near  a multipolar instability, but it also  provides a more accurate dispersion of the single-magnon mode ($m=1$) in comparison with the ordinary SWT. This observation implies that the GSWT is not only necessary for describing multipolar orderings, but also quantitatively more accurate for describing the single-magnon dispersion of usual dipolar orderings whenever the Hamiltonian  includes non-linear on-site spin operators ({\it e.g.}  biquadratic \cite{papanicolaou1988,Batista02,harada2002,Batista04,penc2011,toth2012}  or single-ion anisotropy terms~\cite{Zapf06,Kohama12,wierschem2012}).

\subsection{Excitation spectrum and stability of antiferromagnetic phase}

For  the AFM mean field ground state we have  
${\bm a}=(1, -1,-1 )$. 
By replacing this expression in  Eq.~\eqref{hgsw}, we obtain
\begin{eqnarray}
\mathcal{H}^{\rm AFM}_{\rm GSW} \!\! &=& \!\! \!\!  \sum_{{\bm k}, m}  \mu_{m} b_{{\bm k}m}^{\dagger} b_{{\bm k}m} + 
\Delta_{{\bm k} m} (b^{\dagger} _{{\bm k}m} b^{\dagger}_{-{\bm k}m} + b^{\;} _{{\bm k}m} b^{\;}_{-{\bm k}m}),
\nonumber \\
\end{eqnarray} 
with
  \begin{eqnarray}
    \mu_m &=& 2mdJS \left[ \cos\alpha-\left(2-\frac{2 (m+1)S -m-1}{2S^2}\right)\sin\alpha\right], 
    \nonumber
  \end{eqnarray}
  for $ 1 \leq m \leq 2S$, 

\begin{eqnarray}
\Delta_{{\bm k} 1} &=& JS\left[\cos\alpha-\left(2 -\frac{2S-1}{S^2}\right)\sin\alpha\right] \frac{\gamma_{\bm k}}{2}
\nonumber \\
\Delta_{{\bm k} 2} &=& 2 J \left(1 -\frac{1}{2S}\right) \sin\alpha \frac{\gamma_{\bm k}}{2}
\nonumber
\end{eqnarray}
and $\Delta_{{\bm k} m}=0$ for $ 3 \leq m \leq 2S$.

\begin{table*}
	\begin{tabular}{c|cc|cc}
	\toprule
	$S$&$\alpha_{\rm{min}}^{\rm AF}$ &$m_{\rm soft}$& $\alpha_{\rm{max}}^{\rm AF}$ &$m_{\rm soft}$    
	\\ \colrule
	1                  &$-\frac{\pi}{2}$&2&$\frac{\pi}{4}$&1 and 2\\
	$\frac{3}{2}$ 
	&$\cos \alpha=  -\frac{2}{\sqrt{85}}\cap \sin \alpha= -\frac{9}{\sqrt{85}}$&2 and 3
	&$\cos \alpha=\frac{10}{\sqrt{181}}  \cap \sin \alpha=\frac{9}{\sqrt{181}}  $&1 and 2\\
	$2$
	&$\cos \alpha=  -\frac{1}{\sqrt{65}}\cap \sin \alpha= -\frac{8}{\sqrt{65}}$&4
	&$\cos \alpha=\frac{5}{\sqrt{41}}  \cap \sin \alpha=\frac{4}{\sqrt{41}}  $&1 and 2\\
	$\frac{5}{2}$
	&$\cos \alpha=  -\frac{2}{\sqrt{629}}\cap \sin \alpha= -\frac{25}{\sqrt{629}}$&5
	&$\cos \alpha=\frac{34}{\sqrt{1781}}  \cap \sin \alpha=\frac{25}{\sqrt{1781}}  $&1 and 2\\
	\botrule
	\end{tabular}
	\caption{Range of stability of the AFM phase that is extracted from the analysis of the excitation spectrum.}
	\label{tab2}
\end{table*}

\begin{figure*}[htp]
 \includegraphics[angle=0,width=16cm]{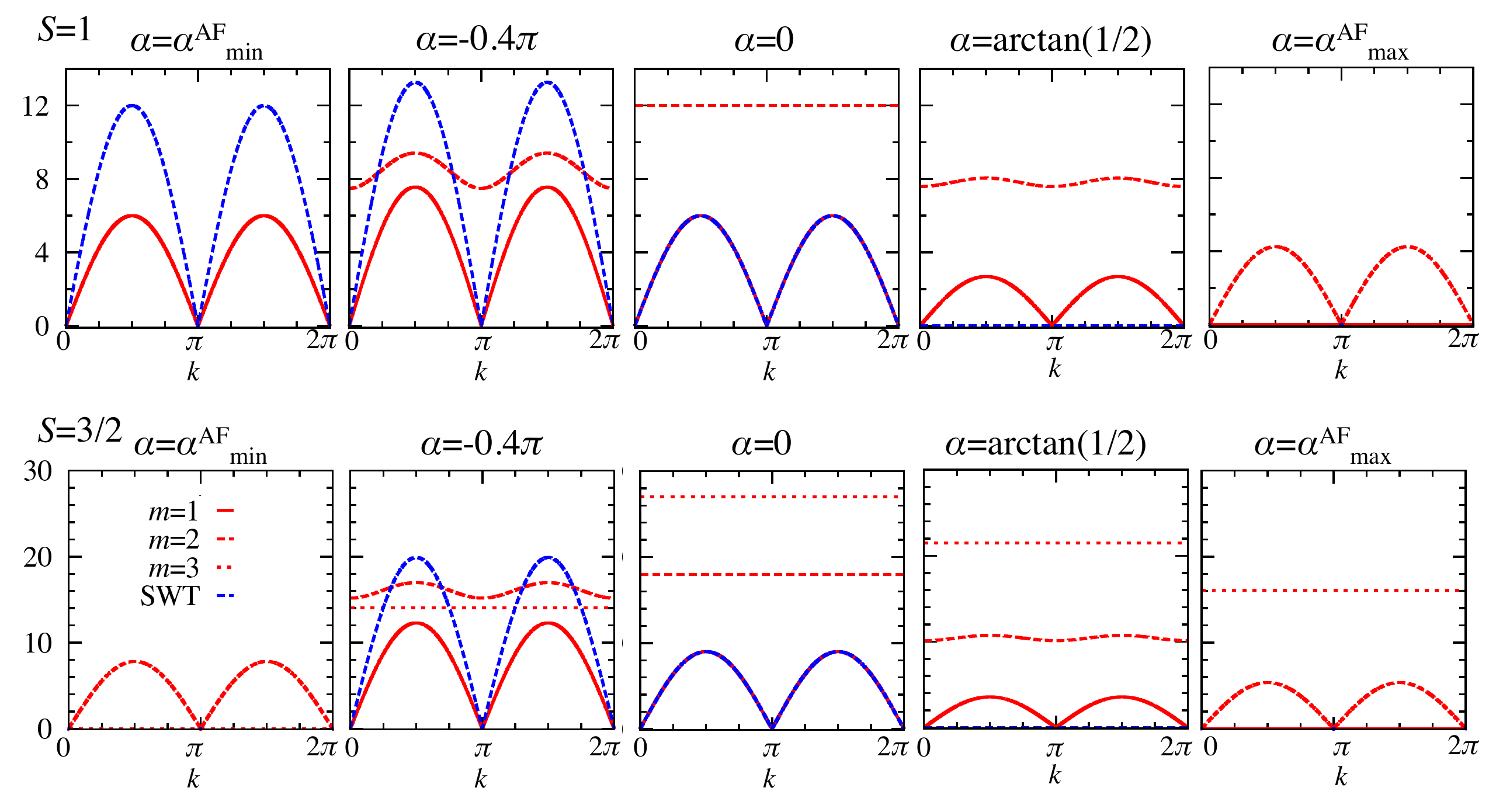}
 \caption{(Color online) 
 	Excitation spectra of the SWT and the GSWT for $S=1$ (upper panels) and $S=3/2$ (lower panels), $\alpha^{AF}_{\rm min} \leq \alpha \leq  \alpha^{AF}_{\rm max}$ and $d=3$. The label $k$ of the horizontal axes corresponds to a parametrization of the wave-vectors along the $(1,1,1)$ direction: ${\bm k}=(k,k,k)$.
 }
 \label{fig:dispAF}
\end{figure*}

By applying a Bogoliubov transformation to the ${b}_{{\bm k}1}$
and ${b}_{{\bm k}2}$ bosons, we obtain
\begin{eqnarray}
\mathcal{H}^{\rm AFM}_{\rm GSW} \!\! &=& \!\! \!\!  \sum_{{\bm k}} (\epsilon^{af}_{{\bm k}1}
\tilde{b}_{{\bm k}1}^{\dag} \tilde{b}_{{\bm k}1} +
\epsilon^{af}_{{\bm k}2}
\tilde{b}_{{\bm k}2}^{\dag} \tilde{b}_{{\bm k}2} )
+ \sum^{2S}_{m=3}  \mu_m \tilde{b}_{{\bm k}m}^{\dag} \tilde{b}_{{\bm k}m}
 \nonumber
\end{eqnarray}
 with 
\begin{eqnarray}
\label{gswafd}
\epsilon^{af}_{{\bm k}1} &=&   
J\left[S\cos\alpha-\left(2S-2+\frac{1}{S}\right)\sin\alpha\right]\sqrt{ 4d^2 -\gamma_{\bm k}^2 }
\nonumber\\
\epsilon^{af}_{{\bm k}2} &=& \sqrt{ 
\mu_2^2 
	- 4 J^2 \left(1-\frac{1}{2S}\right)^{2} \sin^{2}\alpha \gamma_{\bm k}^{2}}.
\end{eqnarray}
Like for the FM case, we note that  the dispersion relation of the single branch of magnon excitations that is obtained with the ordinary SWT,  
\begin{eqnarray}
	\omega_{\bm k}=JS(\cos \alpha - 2 \sin\alpha)\sqrt{ 4d^2-\gamma^2_{\bm k}},
	\label{eq:wkafmswt}
\end{eqnarray}
differs from the single-magnon dispersion, $\epsilon^{af}_{{\bm k}1}$,  predicted by the GSWT. The difference is a global multiplicative factor that goes to one for $J_{Q} \to 0$ or $S \to \infty$, implying
that the spin velocity and the bandwidth of the single-magnon dispersions obtained with the SWT and the GSWT are different for finite $S$ and $J_{Q} \neq 0$ (see Fig.~\ref{fig:dispAF} ).

Fig.~\ref{fig:dispAF} shows the  dispersions for $S=1$ and $S=3/2$. As for the FM case, the dispersion relations obtained with the GSWT imply the following  stability conditions for the AFM ordering:
\begin{eqnarray}
S\cos\alpha-\left(2S-2+\frac{1}{S}\right)\sin\alpha &>& 0 , \nonumber\\
S\cos\alpha-\left(2S-3+\frac{3}{2S}\right)\sin\alpha - \left(1-\frac{1}{2S}\right)\left|\sin\alpha\right| & >&0 , 
\nonumber\\
S\cos\alpha-\left(2S- m-1+\frac{m+1}{2S}\right)\sin\alpha & >&0 , \nonumber \\
\label{condafm}
\end{eqnarray}
for $ 3 \leq m \leq 2S$. Table~\ref{tab2} summarizes the expected phase boundaries and the multipolar order, $m$, of the mode that becomes  soft  for $S=1 \sim 5/2$. We reiterate that the actual region of stability can be smaller than the one obtained by this analysis if
the transition to a different phase is of first order. 

Fig.~\ref{fig:dispAF} also shows the single-magnon dispersion that results from ordinary SWT. The situation is analogous to the FM case. Direct comparison of Eqs.~\eqref{gswafd} and \eqref{eq:wkafmswt} shows that the prefactors of the contributions proportional to $\sin{\alpha}$ differ by 
$(1-1/S+1/2S^2)$. This difference has its roots in the same biquadratic terms that we already discussed for the FM case. Again, SWT predicts instabilities at 
$\alpha={\rm atan}(1/2)$ which are very far from the actual phase boundaries. In addition, the single-magnon dispersion predicted by SWT for $\alpha<0$ is much broader than the one obtained with the GSWT. The overall conclusion is the same as for the FM  case: the GSWT gives a quantitatively  more accurate estimate of the single-magnon dispersion of usual dipolar orderings when the Hamiltonian contains non-linear on-site spin operators.

\subsection{Phase boundaries}

\begin{figure}[htp!]
 \includegraphics[angle=0,width=7.2cm]{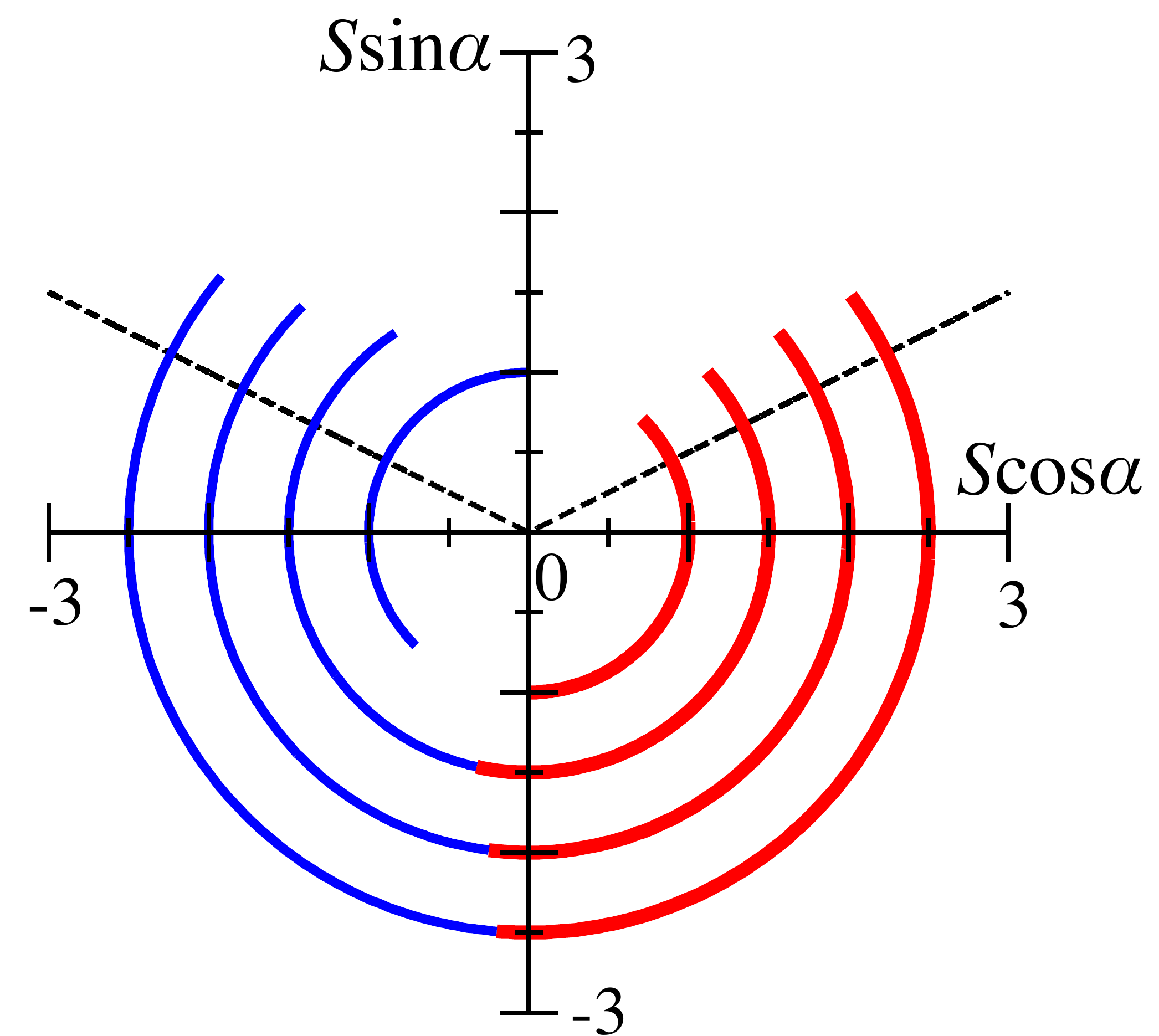}
 \caption{(Color online) 
 	Regions of stability of the FM (Blue) and AFM (Red) orderings in the generalized spin-wave theory. Radii corresponds to $S$.
 	The dashed line (Black) shows where these magnetic orderings become unstable in the ordinary spin-wave theory. 
 }
 \label{fig:fig1}
 \end{figure}

Fig.~\ref{fig:fig1} shows the regions of stability of the FM and AFM orderings obtained from the excitation spectra predicted by the GSWT.
Several studies of ${\cal H}$ have been reported for $S=1$ spins on square lattices \cite{papanicolaou1988,Batista02,harada2002,Batista04,penc2011,blume1969,toth2012}. 
In particular, the boundaries of the FM phase obtained from world-line Monte-Carlo (WLMC) simulations \cite{harada2002}  (there is no negative sign problem for $J_Q<0$) are identical to those obtained with the GSWT .
Moreover, the WLMC reveals that a ferroquadrupolar phase is stabilized in the region $5\pi/4<\alpha<3\pi/2$ in
agreement with the result obtained with the GSWT: the  $m=2$ and ${\bm k}={\bm 0}$ mode becomes soft at $\alpha=5\pi/4$.
A combination of analytical and numerical results~\cite{toth2012}  predicts that
 a three-sublattice quadrupolar  and a three-sublattice magnetic phase are stabilized in the regions $\pi/2>\alpha>\pi/4$
and $\pi/4>\alpha > \sim 0.2\pi $, respectively. We note that ${\cal H}$ is invariant under global SU(3) transformations for $\alpha = \pm \pi/4$, implying that the dipolar and nematic order parameters must coexist at these two points with the same ordering wave-vector because they are connected by a global symmetry operation of ${\cal H}$ \cite{Batista02,Batista04}. The three sublattice structures obtained for $\pi/4>\alpha >  \sim 0.2 \pi $ are also obtained within the GSWT after  a proper minimization of the mean field energy as a function of $\alpha$ (not included here). The mean field ground state is largely degenerate in this interval and the optimal ordering is selected by quantum fluctuations via the order from disorder mechanism. In other words, it is necessary to compute  
$H_{GSWT}$ for all the degenerate product states \eqref{min} that minimize the mean field energy,  and determine the one that 
minimizes the ground state energy of $H_{GSWT}$. Because the obtained three-sublattice structure and the AFM phase have 
different ordering wave-vectors, it is not possible to determine the transition between both phase by analysing the 
spectrum of excitations of the AFM state. Note that  according to such analysis, the range of stability of the 
AFM phase could extend up to $\alpha = \pi/4$ (see Table~\ref{tab2}).

For $S>1$, the analysis of the excitation spectrum  obtained with the GSWT  suggest the possibility of a  continuous quantum phase transition from the FM phase to a new phase with a soft $2S$ mode
for a positive critical value of $J_Q$ and $J_L=-1$ (see Table~\ref{tab1}). The corresponding
critical value of $\alpha=\alpha^{\rm F}_{\rm min}$ is then determined by the $m=1,2$ inequalities  listed 
in Eq.~\eqref{condfm}:
\begin{eqnarray}
\label{alfminF}
  \tan \alpha_{\rm min}^{\rm F} =-\frac{S}{2(S-1)}.
\end{eqnarray}
We note that the mean field ordering that results from  minimization of  $\langle \psi_{\rm mf} | {\cal H} |\psi_{\rm mf} \rangle$ 
(see Eq.~\ref{min})  over two and three-sublattice structures leads to first order transitions exactly at the values of $\alpha_{\rm min}^{\rm F} $ and $\alpha_{\rm max}^{\rm F}$ that are  listed in Table~\ref{tab1}. In the new ordered mean field state for $\alpha \lesssim \alpha_{\rm min}^{\rm F}$, the spins
of one sublattice remain fully polarized, while the spins of the other sublattice become only partially polarized:
${\tilde b}^{\dagger}_{{\bm r }0} | \emptyset \rangle = | S^z=S \rangle$ for ${\bm r }$ in the A sublattice and 
${\tilde b}^{\dagger}_{{\bm r }0}  | \emptyset \rangle = | S^z=S-1 \rangle$ for ${\bm r }$ in the B sublattice. Such a ferrimagnetic  mean field state 
has no classical counterpart at $T=0$ because classical spins are always fully polarized along a particular direction. In other words, the spins of the B sublattice have a nematic component which is larger than the side effect produced by usual magnetic ordering. 
To quantify this statement we simply note that $\langle {\bm S}_{\bm r} \cdot {\bm S}_{\bm r}  \rangle - \langle {\bm S}_{\bm r} \rangle  \cdot  \langle {\bm S}_{\bm r}  \rangle $ is equal to $S$ for spins on the A sublattice and to $3S-1$ for spins on the B sublattice. The anomalous variance of spins in the B sublattice is caused by a nematic component which is larger than the one obtained for fully polarized spins. 

The same mean field analysis indicates a direct first order transition between the FM and AFM 
phases exactly at the $\alpha=\alpha_{\rm max}^{\rm F}=\alpha_{\rm min}^{\rm AF}$ values that are listed in Tables~\ref{tab1} and \ref{tab2}.
 They are determined from the $m=2S$ inequalities  listed  in Eq.~(\ref{condfm},\ref{condafm}):
\begin{eqnarray}
\label{alfamaxFminAF}
\tan \alpha_{\rm max}^{\rm F} = \tan \alpha_{\rm min}^{\rm AF} =2 S^2 ,
\end{eqnarray}
for $S = \frac{3}{2}$ the mode $m = 2$ also becomes soft at the transition.

A similar analysis suggests that the AFM phase may undergo a transition to a  phase with quadrupolar and/or another 
dipolar ordering for $J_L=1$ and a positive critical value of $J_Q$, i.e., the $m=1,2$ modes become soft at the same 
critical value of $\alpha = \alpha_{\rm max}^{\rm AF} $. Therefore,  $\alpha^{\rm AF}_{\rm max}$ is determined
from the $m=1,2$ inequalities  listed  in Eq.~\eqref{condafm}:
\begin{eqnarray}
\label{alfamaxAF}
\tan \alpha_{\rm max}^{\rm AF} =\frac{S^2}{2S^2-2S+1} .
\end{eqnarray}
Similarly to the FM case, minimization of $\langle \psi_{\rm mf} | {\cal H} |\psi_{\rm mf} \rangle$ 
 (see Eq.~\ref{min}) over two and three-sublattice structures leads to a first order transition into an AFM state
in which the spins remain fully polarized in one sublattice and become only partially polarized in the other sublattice:
${\tilde b}^{\dagger}_{{\bm r }0} | \emptyset \rangle= | S^z=S \rangle$ for ${\bm r }$ in the A sublattice and 
${\tilde b}^{\dagger}_{{\bm r }0} | \emptyset \rangle= | S^z=-S+1 \rangle$ for ${\bm r }$ in the B sublattice.
Again, the spins of the B sublattice have an anomalously large nematic component that is a pure quantum effect. 
The first order mean field transition again occurs exactly at the critical values of $\alpha^{AF}_{\rm max}$ given in Eq.~\eqref{alfamaxAF} and listed in Table~\ref{tab2}.

Once again, we emphasize that $\alpha_{\rm min}^{\rm F} \to \pi - \arctan{1/2}$  and $\alpha_{\rm max}^{\rm AF} \to \arctan{1/2}$ for $S \to \infty$, implying that the instabilities obtained with the SWT (see Figs.~\ref{fig:dispF} and \ref{fig:dispAF}) coincide with the ones predicted by the GSWT {\it only in the} $S \to \infty$ {\it limit}. For finite $S$, the SWT predicts smaller stability ranges of the FM and AFM phases than the ones obtained from energy minimization over two and three-sublattice product or mean field states (Eq.~\ref{min}).  We also note that the nematic instabilities at $\alpha^{F}_{\rm min}$ and $\alpha^{AF}_{\rm max}$ (see Tables~\ref{tab1} and \ref{tab2})  survive in the  $S \to \infty$ limit, suggesting that the phase diagram for $S \to \infty$ does not coincide with the one obtained for a classical vector field.  This simple result indicates that the large $S$ limit is not  necessarily  equal to the classical spin limit.


\section{Conclusions}

The results of the previous sections clearly illustrate the advantages of using the GSWT instead of the ordinary SWT. Besides the obvious case of non-dipolar $T=0$ orderings, which do not have a classical counterpart, we have shown that, even for dipolar orderings, the GSWT leads to a low-energy spectrum of excitations that is quantitatively and qualitatively better than the spectrum predicted by SWT. In the first place, the  spectrum of the GSWT contains multipolar modes, in addition to the single-magnon modes of SWT, that become part of the low-energy spectrum near quantum phase transitions. It is clear that these additional low-energy modes give a substantial contribution to the low-temperature thermodynamic properties near quantum critical points. Moreover, the GSWT predicts a more accurate single-magnon dispersion whenever the Hamiltonian includes on-site operators, which are non-linear in the spin components ({\it e.g.}  biquadratic \cite{papanicolaou1988,Batista02,harada2002,Batista04,penc2011,toth2012}  or single-ion anisotropy terms~\cite{Zapf06,Kohama12,wierschem2012}). The presence of these terms can lead to important differences in the spin-wave velocities predicted by both theories, as well as in the stability range of  dipolar phases. Finally, the inclusion of multi-polar modes $m>1$ in the GSWT allows to detect multipolar instabilities already at the linear level, {\it i.e.}, without including interactions between modes. For instance, the $m=3$ soft mode that we found for $\alpha=\alpha^{AF}_{\rm min}$ and $S=3/2$ indicates that 
a small bicubic term should be enough to induce octupolar ordering around $\alpha=\alpha^{AF}_{\rm min}$. 

The GSWT is an expansion in powers of  $1/\sqrt{N {\cal S} }$.  To go beyond the linear order (include interaction between modes), it is convenient not to replace the  $N {\cal S}$  by one in Eq.\eqref{eq:constr} in order to keep track of the  $1/N {\cal S}$  order of each diagram. Like in the case of ordinary SWT, an order by order expansion in  $1/N {\cal S}$  preserves Goldstone modes associated with spontaneously broken continuous symmetries. 

An alternative approach to the Holstein-Primakoff approximation is to introduce a Lagrange multiplier which enforces the constraint \eqref{eq:constr} and allow the condensate fraction to take a value, $\langle  {\tilde b}_{{\bm r}0}\rangle=s$, which is obtained by minimization of the ground state energy \cite{wierschem2012}. This is not an order by order  $1/ N {\cal S} $ expansion. Consequently, this approach does not preserve the Goldstone modes associated with spontaneously broken continuous symmetries. However, it may still be very appropriate for describing gapped phases that do not break any continuous symmetry \cite{wierschem2012}. Indeed, the bond operators introduced by  Sachdev and Bahtt \cite{Sachdev90} are a particular example of this approach. The four bosons that create the singlet state and the three triplets of a single-dimer (natural unit cell of dimerized lattices) can be identified with the four SBs associated with the fundamental representation of SU(4). 

Finally, we would like to mention that although multi-flavored bosons have been used several times in the past to attack spin problems (see for instance \cite{papanicolaou1988,Sachdev90,toth2012}), we are not aware of any attempt to provide a geometric interpretation of such approaches, or relate them to preexisting formalisms. In the present manuscript we are doing both things by demonstrating that these approaches correspond to a generalization of the usual SWT from SU(2) to SU(N). Moreover, we are also demonstrating that under quite general conditions this GSWT is better than the usual SWT even for describing dipolar orderings.










\begin{acknowledgements}
Work at the LANL was performed under the auspices of the U.S.\ DOE contract No.~DE-AC52-06NA25396 through the LDRD program. This material is based upon work supported in part by the NSF under Grant No. 
PHY-1066293 and the hospitality of the Aspen Center for Physics. 
R. Muniz also thanks CNPq (Brazil) for financial support. 
\end{acknowledgements}

\bibliographystyle{apsrev}
\bibliography{gsw}

\begin{thebibliography}{18}
\expandafter\ifx\csname natexlab\endcsname\relax\def\natexlab#1{#1}\fi
\expandafter\ifx\csname bibnamefont\endcsname\relax
  \def\bibnamefont#1{#1}\fi
\expandafter\ifx\csname bibfnamefont\endcsname\relax
  \def\bibfnamefont#1{#1}\fi
\expandafter\ifx\csname citenamefont\endcsname\relax
  \def\citenamefont#1{#1}\fi
\expandafter\ifx\csname url\endcsname\relax
  \def\url#1{\texttt{#1}}\fi
\expandafter\ifx\csname urlprefix\endcsname\relax\def\urlprefix{URL }\fi
\providecommand{\bibinfo}[2]{#2}
\providecommand{\eprint}[2][]{\url{#2}}

\bibitem[{\citenamefont{Shiina et~al.}(1997)\citenamefont{Shiina, Shiba, and
  Thalmeier}}]{shiina1997}
\bibinfo{author}{\bibfnamefont{R.}~\bibnamefont{Shiina}},
  \bibinfo{author}{\bibfnamefont{H.}~\bibnamefont{Shiba}}, \bibnamefont{and}
  \bibinfo{author}{\bibfnamefont{P.}~\bibnamefont{Thalmeier}},
  \bibinfo{journal}{J. Phys. Soc. Jpn.} \textbf{\bibinfo{volume}{66}},
  \bibinfo{pages}{1741} (\bibinfo{year}{1997}).

\bibitem[{\citenamefont{Sakai et~al.}(1997)\citenamefont{Sakai, Shiina, Shiba,
  and Thalmeier}}]{sakai1997}
\bibinfo{author}{\bibfnamefont{O.}~\bibnamefont{Sakai}},
  \bibinfo{author}{\bibfnamefont{R.}~\bibnamefont{Shiina}},
  \bibinfo{author}{\bibfnamefont{H.}~\bibnamefont{Shiba}}, \bibnamefont{and}
  \bibinfo{author}{\bibfnamefont{P.}~\bibnamefont{Thalmeier}},
  \bibinfo{journal}{J. Phys. Soc. Jpn.} \textbf{\bibinfo{volume}{66}},
  \bibinfo{pages}{3005} (\bibinfo{year}{1997}).

\bibitem[{\citenamefont{Ray and Sivardiere}(1976)}]{ray1976}
\bibinfo{author}{\bibfnamefont{D.}~\bibnamefont{Ray}} \bibnamefont{and}
  \bibinfo{author}{\bibfnamefont{J.}~\bibnamefont{Sivardiere}},
  \bibinfo{journal}{Solid State Communications} \textbf{\bibinfo{volume}{19}},
  \bibinfo{pages}{1053} (\bibinfo{year}{1976}).

\bibitem[{\citenamefont{Tsunetsugu and Arikawa}(2006)}]{tsunetsugu2006}
\bibinfo{author}{\bibfnamefont{H.}~\bibnamefont{Tsunetsugu}} \bibnamefont{and}
  \bibinfo{author}{\bibfnamefont{M.}~\bibnamefont{Arikawa}},
  \bibinfo{journal}{J. Phys. Soc. Jpn.} \textbf{\bibinfo{volume}{75}},
  \bibinfo{pages}{083701} (\bibinfo{year}{2006}).

\bibitem[{\citenamefont{L{\"a}uchli et~al.}(2006)\citenamefont{L{\"a}uchli,
  Mila, and Penc}}]{lauchli2006}
\bibinfo{author}{\bibfnamefont{A.}~\bibnamefont{L{\"a}uchli}},
  \bibinfo{author}{\bibfnamefont{F.}~\bibnamefont{Mila}}, \bibnamefont{and}
  \bibinfo{author}{\bibfnamefont{K.}~\bibnamefont{Penc}},
  \bibinfo{journal}{Physical review letters} \textbf{\bibinfo{volume}{97}},
  \bibinfo{pages}{87205} (\bibinfo{year}{2006}).

\bibitem[{\citenamefont{Papanicolaou}(1988)}]{papanicolaou1988}
\bibinfo{author}{\bibfnamefont{N.}~\bibnamefont{Papanicolaou}},
  \bibinfo{journal}{Nuclear Physics B} \textbf{\bibinfo{volume}{305}},
  \bibinfo{pages}{367} (\bibinfo{year}{1988}).

\bibitem[{\citenamefont{Auerbach}(1998)}]{auerbach98}
\bibinfo{author}{\bibfnamefont{A.}~\bibnamefont{Auerbach}},
  \emph{\bibinfo{title}{Interacting Electrons and Quantum Magnetism}}
  (\bibinfo{publisher}{Springer}, \bibinfo{year}{1998}).

\bibitem[{\citenamefont{Batista et~al.}(2002)\citenamefont{Batista, Ortiz, and
  Gubernatis}}]{Batista02}
\bibinfo{author}{\bibfnamefont{C.~D.} \bibnamefont{Batista}},
  \bibinfo{author}{\bibfnamefont{G.}~\bibnamefont{Ortiz}}, \bibnamefont{and}
  \bibinfo{author}{\bibfnamefont{J.~E.} \bibnamefont{Gubernatis}},
  \bibinfo{journal}{Phys. Rev. B} \textbf{\bibinfo{volume}{65}},
  \bibinfo{pages}{180402} (\bibinfo{year}{2002}).

\bibitem[{\citenamefont{Batista and Ortiz}(2004)}]{Batista04}
\bibinfo{author}{\bibfnamefont{C.~D.} \bibnamefont{Batista}} \bibnamefont{and}
  \bibinfo{author}{\bibfnamefont{G.}~\bibnamefont{Ortiz}},
  \bibinfo{journal}{Advances in Physics} \textbf{\bibinfo{volume}{53}},
  \bibinfo{pages}{1} (\bibinfo{year}{2004}).

\bibitem[{\citenamefont{Smerald and Shannon}(2013)}]{Smerald13}
\bibinfo{author}{\bibfnamefont{A.}~\bibnamefont{Smerald}} \bibnamefont{and}
  \bibinfo{author}{\bibfnamefont{N.}~\bibnamefont{Shannon}},
  \bibinfo{journal}{arXiv:1303.4465 [cond-mat.str-el]}  (\bibinfo{year}{2013}).

\bibitem[{\citenamefont{Zapf et~al.}(2006)\citenamefont{Zapf, Zocco, Hansen,
  Jaime, Harrison, Batista, Kenzelmann, Niedermayer, Lacerda, and
  Paduan-Filho}}]{Zapf06}
\bibinfo{author}{\bibfnamefont{V.~S.} \bibnamefont{Zapf}},
  \bibinfo{author}{\bibfnamefont{D.}~\bibnamefont{Zocco}},
  \bibinfo{author}{\bibfnamefont{B.~R.} \bibnamefont{Hansen}},
  \bibinfo{author}{\bibfnamefont{M.}~\bibnamefont{Jaime}},
  \bibinfo{author}{\bibfnamefont{N.}~\bibnamefont{Harrison}},
  \bibinfo{author}{\bibfnamefont{C.~D.} \bibnamefont{Batista}},
  \bibinfo{author}{\bibfnamefont{M.}~\bibnamefont{Kenzelmann}},
  \bibinfo{author}{\bibfnamefont{C.}~\bibnamefont{Niedermayer}},
  \bibinfo{author}{\bibfnamefont{A.}~\bibnamefont{Lacerda}}, \bibnamefont{and}
  \bibinfo{author}{\bibfnamefont{A.}~\bibnamefont{Paduan-Filho}},
  \bibinfo{journal}{Phys. Rev. Lett.} \textbf{\bibinfo{volume}{96}},
  \bibinfo{pages}{077204} (\bibinfo{year}{2006}).

\bibitem[{\citenamefont{Penc and L{\"a}uchli}(2011)}]{penc2011}
\bibinfo{author}{\bibfnamefont{K.}~\bibnamefont{Penc}} \bibnamefont{and}
  \bibinfo{author}{\bibfnamefont{A.~M.} \bibnamefont{L{\"a}uchli}},
  \emph{\bibinfo{title}{Introduction to Frustrated Magnetism}}
  (\bibinfo{publisher}{Springer}, \bibinfo{year}{2011}),
  chap.~\bibinfo{chapter}{13}, pp. \bibinfo{pages}{331--362}.

\bibitem[{\citenamefont{Wierschem et~al.}(2012)\citenamefont{Wierschem, Kato,
  Nishida, Batista, and Sengupta}}]{wierschem2012}
\bibinfo{author}{\bibfnamefont{K.}~\bibnamefont{Wierschem}},
  \bibinfo{author}{\bibfnamefont{Y.}~\bibnamefont{Kato}},
  \bibinfo{author}{\bibfnamefont{Y.}~\bibnamefont{Nishida}},
  \bibinfo{author}{\bibfnamefont{C.~D.} \bibnamefont{Batista}},
  \bibnamefont{and} \bibinfo{author}{\bibfnamefont{P.}~\bibnamefont{Sengupta}},
  \bibinfo{journal}{Phys. Rev. B} \textbf{\bibinfo{volume}{86}},
  \bibinfo{pages}{201108} (\bibinfo{year}{2012}).

\bibitem[{\citenamefont{Blume and Hsieh}(1969)}]{blume1969}
\bibinfo{author}{\bibfnamefont{M.}~\bibnamefont{Blume}} \bibnamefont{and}
  \bibinfo{author}{\bibfnamefont{Y.}~\bibnamefont{Hsieh}}, \bibinfo{journal}{J.
  Appl. Phys.} \textbf{\bibinfo{volume}{40}}, \bibinfo{pages}{1249}
  (\bibinfo{year}{1969}).

\bibitem[{\citenamefont{Harada and Kawashima}(2002)}]{harada2002}
\bibinfo{author}{\bibfnamefont{K.}~\bibnamefont{Harada}} \bibnamefont{and}
  \bibinfo{author}{\bibfnamefont{N.}~\bibnamefont{Kawashima}},
  \bibinfo{journal}{Phys. Rev. B} \textbf{\bibinfo{volume}{65}},
  \bibinfo{pages}{052403} (\bibinfo{year}{2002}).

\bibitem[{\citenamefont{T{\'o}th et~al.}(2012)\citenamefont{T{\'o}th,
  L{\"a}uchli, Mila, and Penc}}]{toth2012}
\bibinfo{author}{\bibfnamefont{T.~A.} \bibnamefont{T{\'o}th}},
  \bibinfo{author}{\bibfnamefont{A.~M.} \bibnamefont{L{\"a}uchli}},
  \bibinfo{author}{\bibfnamefont{F.}~\bibnamefont{Mila}}, \bibnamefont{and}
  \bibinfo{author}{\bibfnamefont{K.}~\bibnamefont{Penc}},
  \bibinfo{journal}{Phys. Rev. B} \textbf{\bibinfo{volume}{85}},
  \bibinfo{pages}{140403} (\bibinfo{year}{2012}).

\bibitem[{\citenamefont{Kohama et~al.}(2011)\citenamefont{Kohama, Sologubenko,
  Dilley, Zapf, Jaime, Mydosh, Paduan-Filho, Al-Hassanieh, Sengupta,
  Gangadharaiah et~al.}}]{Kohama12}
\bibinfo{author}{\bibfnamefont{Y.}~\bibnamefont{Kohama}},
  \bibinfo{author}{\bibfnamefont{A.~V.} \bibnamefont{Sologubenko}},
  \bibinfo{author}{\bibfnamefont{N.~R.} \bibnamefont{Dilley}},
  \bibinfo{author}{\bibfnamefont{V.~S.} \bibnamefont{Zapf}},
  \bibinfo{author}{\bibfnamefont{M.}~\bibnamefont{Jaime}},
  \bibinfo{author}{\bibfnamefont{J.~A.} \bibnamefont{Mydosh}},
  \bibinfo{author}{\bibfnamefont{A.}~\bibnamefont{Paduan-Filho}},
  \bibinfo{author}{\bibfnamefont{K.~A.} \bibnamefont{Al-Hassanieh}},
  \bibinfo{author}{\bibfnamefont{P.}~\bibnamefont{Sengupta}},
  \bibinfo{author}{\bibfnamefont{S.}~\bibnamefont{Gangadharaiah}},
  \bibnamefont{et~al.}, \bibinfo{journal}{Phys. Rev. Lett.}
  \textbf{\bibinfo{volume}{106}}, \bibinfo{pages}{037203}
  (\bibinfo{year}{2011}).

\bibitem[{\citenamefont{Sachdev and Bhatt}(1990)}]{Sachdev90}
\bibinfo{author}{\bibfnamefont{S.}~\bibnamefont{Sachdev}} \bibnamefont{and}
  \bibinfo{author}{\bibfnamefont{R.~N.} \bibnamefont{Bhatt}},
  \bibinfo{journal}{Phys. Rev. B} \textbf{\bibinfo{volume}{41}},
  \bibinfo{pages}{9323} (\bibinfo{year}{1990}).

\end{thebibliography}

\end{document}